\documentclass[aip, amsmath, amssymb, reprint]{revtex4-1}

\usepackage{algorithmicx}
\usepackage{algpseudocode}
\usepackage{amsmath}
\usepackage{amssymb}
\usepackage{array}
\usepackage{color}
\usepackage{dcolumn}
\usepackage[T1]{fontenc}
\usepackage{float}
\usepackage{hyperref}
\usepackage{graphicx}
\usepackage[utf8]{inputenc}
\usepackage{lmodern}
\usepackage{makecell}
\usepackage[version=4]{mhchem}
\usepackage{physics}
\usepackage[normalem]{ulem}
\usepackage[dvipsnames]{xcolor}
\usepackage{xspace}
\usepackage{xfrac}

\newcommand{\abinitio}{\textit{ab initio}\xspace}

\newcommand{\etal}{\textit{et al.}\xspace}

\newcommand{\ang}{\text{\,\AA}\xspace}

\newcommand{\gpa}{\text{\,GPa}\xspace}
\newcommand{\icubang}{\text{\,\AA$^{-3}$}\xspace}
\newcommand{\kel}{\text{\,K}\xspace}
\newcommand{\wvn}{\text{\,cm$^{-1}$}\xspace}
\newcommand{\sh}{\text{-}}

\newcommand{\hcp}{\textit{hcp}\xspace}

\newcommand{\parahyd}{\textit{para}-{H}$_2$\xspace}
\newcommand{\para}{\textit{para}-H$_2$\xspace}

\newcommand{\sampT}{\text{S}$[2]$\xspace}        % sampling-two
\newcommand{\sampTT}{\text{S}$[2,3]$\xspace}     % sampling-two-three
\newcommand{\sampTTF}{\text{S}$[2,3,4]$\xspace}  % sampling-two-three-four

\newcommand{\estiT}{\text{E}$[2]$\xspace}        % estimating-two
\newcommand{\estiTT}{\text{E}$[2,3]$\xspace}     % estimating-two-three
\newcommand{\estiTTF}{\text{E}$[2,3,4]$\xspace}  % estimating-two-three-four

\usepackage{silence}
\begin{document}

%%%%%%%%%%%%%%%%%%%%%%%%%%%%%%%%%%%%%%%%%%%%
%%% TITLE, AUTHORSHIP, EMAIL, AFFILIATIONS
%%%%%%%%%%%%%%%%%%%%%%%%%%%%%%%%%%%%%%%%%%%%

\title{
    Path-integral Monte Carlo simulations of solid parahydrogen
    using two-body, three-body, and four-body
    \textit{ab initio} interaction potential energy surfaces
}

\author{Alexander Ibrahim}
\affiliation{
    Department of Physics and Astronomy,
    University of Waterloo,
    200 University Avenue West, Waterloo, Ontario N2L 3G1, Canada
}
\affiliation{
    Department of Chemistry,
    University of Waterloo,
    200 University Avenue West, Waterloo, Ontario N2L 3G1, Canada
}

\author{Pierre-Nicholas Roy}
\email{pnroy@uwaterloo.ca}
\affiliation{
    Department of Chemistry,
    University of Waterloo,
    200 University Avenue West, Waterloo, Ontario N2L 3G1, Canada
}

%%%%%%%%%%%%%%%%%%%%%%%%%%%%%%%%%%%%%%%%%%%%
%%% ABSTRACT
%%%%%%%%%%%%%%%%%%%%%%%%%%%%%%%%%%%%%%%%%%%%

\begin{abstract}

We present path integral Monte Carlo simulation results for the equation of state of solid parahydrogen
between $ 0.024 \icubang $ and $ 0.1 \icubang $ at $ T = 4.2 \kel $.
The simulations are performed using non-additive isotropic \abinitio
two-body, three-body, and four-body
potential energy surfaces (PES).
We apply corrections to account for both the
finite size simulation errors
and the Trotter factorization errors.
Simulations that use only the two-body PES during sampling
yield an equation of state similar to that of simulations
that use both the two-body and three-body PESs during sampling.
With the four-body interaction energy,
we predict an equilibrium density of $ 0.02608 \icubang $,
very close to the experimental result of $ 0.0261 \icubang $.
The inclusion of the four-body interaction energy
also brings the simulation results in excellent agreement
with the experimental pressure-density data
until around $ 0.065 \icubang $,
beyond which the simulation results overestimate the pressure.
These PESs overestimate the average kinetic energy per molecule
at the equilibrium density by about $ 7 \% $
compared to the experimental result.
Our findings suggest that,
at higher densities,
we require five-body and higher-order many-body interactions
to quantitatively improve the agreement between the pressure-density curve
produced by simulations,
and that of experiment.
Using the four-body PES during sampling
at excessively high densities,
where such higher-order many-body interactions are likely to be significant,
causes an artificial symmetry breaking in the \hcp lattice structure of the solid.

\end{abstract}

%%%%%%%%%%%%%%%%%%%%%%%%%%%%%%%%%%%%%%%%%%%%
%%% TITLE CREATION
%%%%%%%%%%%%%%%%%%%%%%%%%%%%%%%%%%%%%%%%%%%%

\maketitle

%%%%%%%%%%%%%%%%%%%%%%%%%%%%%%%%%%%%%%%%%%%%
%%% INTRODUCTION
%%%%%%%%%%%%%%%%%%%%%%%%%%%%%%%%%%%%%%%%%%%%

\section{Introduction} \label{sec:intro}

Solid parahydrogen (\parahyd)
is a quantum molecular solid
that displays a wide variety of exotic quantum behaviours.
\cite{ph2solidexp:60gush,ph2solidtheo:66nosa,ph2solidexp:80silv,ph2solidexp:21bhan}
The most striking of these behaviours
is its abnormally large zero-point kinetic energy,
which is due to
the low mass of,
and weak intermolecular interactions between,
the \parahyd molecules.
\cite{ph2solidexp:66barr,ph2solidexp:70schn,ph2solidexp:80silv}
Each \parahyd molecule
has a very large translational displacement
about its nominal lattice site.
For example,
at the solid's equilibrium density of $ \rho_0 = 0.0261 \icubang $,
\cite{ph2solidtheo:17duss,ph2solidexp:23pris}
the ratio of
the root mean square displacement of each \parahyd molecule
from its nominal lattice site
to the lattice constant
is about $ 20 \, \% $.
\cite{ph2solidexp:23pris,ph2solidexp:12fern,ph2solidtheo:17duss}
Solid \parahyd can be visualized
as an \hcp lattice
with a nearly spherically symmetric decaying probability cloud
for each \parahyd molecule
centred at each lattice site.

Most simulations of solid parahydrogen
are done using pointwise pair potentials.
\cite{h2pes:83buck,h2pes:78silv,h2pes:84norm,h2pes:00diep,h2pes:08hind,h2pes:08patk,ph2cluster:14faru}
All pair potentials between hydrogen molecules share several common features.
Each one has a strong repulsive wall at short pair distances,
a deep attractive well around $ 3 \ang $,
and a weak decaying attractive tail to model the dispersion interaction.
Pointwise pair potentials are very common because
they are computationally efficient,
easy to implement,
and provide accurate results under certain conditions.
However,
it has become increasingly apparent that
we cannot accurately model solid \parahyd
using only pair potentials,
especially at high densities.
\cite{h2pes:13omiy,ph2solidtheo:06oper}
A complete model of solid \parahyd
requires the use of many-body interactions.

Non-additive many-body interactions,
such as the three-body and four-body interaction,
are very weak compared to the two-body interaction
except at very short intermolecular distances.
\cite{manybody:94elro,manybody:20oter}
In many chemical systems,
many-body interactions are either completely negligible
or can be treated as a slight perturbation to the dominant two-body interaction.
However,
the large translational displacement of each \parahyd molecule
about its nominal lattice site
causes its probability distribution
to spill into very short intermolecular spacings.
As a result,
solid \parahyd is very sensitive to the effects of
non-additive many-body interaction potentials.

Several computational studies have emphasized the importance
of the three-body interaction potential for \parahyd.
\cite{threebody:96wind,threebody:08hind,threebody:10manz,pathinteg:22ibra,virial:13cenc}
The importance of many-body interaction potentials
has also been demonstrated in other systems,
such as
bulk systems of helium
\cite{threebody:00rosc,manybody:06tian,pathinteg:17barn_1,virial:24garb,virial:24bino,manybody:24mari}
solid neon,
\cite{threebody:00rosc,threebody:09schw}
solid and gaseous argon,
\cite{threebody:00rosc,virial:13garb,threebody:24malt}
and
solid krypton.
\cite{fourbody:12tian}
In a recent simulation study,
Wheatley \etal
created an \abinitio four-body potential energy surface (PES) for helium
and investigated its importance in predicting the fourth virial coefficient of helium.
\cite{fourbody:23whea}
In another recent work,
Morresi and Garberoglio
used the aforementioned four-body PES for helium
to perform PIMC simulations of superfluid and normal liquid helium-4.
\cite{manybody:25morr}

Advancements in both quantum chemistry and machine learning
have made it possible to create PESs for \abinitio many-body systems.
\cite{pytorch:19pasz,tensorflow:15abad,gaussianprocesses:22grah}
We collect the \abinitio training data energies by directly solving
the time-independent Schr{\"o}dinger equation on a computer,
and then use a machine learning method
to create a PES that maps the relevant coordinates to their corresponding interaction energies.
\cite{gaussianprocesses:21broa}
Unlike phenomenological PESs,
\abinitio PESs have no dependence on the results of any specific experiment,
and can be constructed to contain the different many-body contributions explicitly.

In this work,
we calculate the equation of state (EOS) of solid parahydrogen
at $ T = 4.2 \kel $ using path-integral Monte Carlo (PIMC) simulations.
The solid is modelled as a hexagonal close-packed (\hcp) lattice
between the densities of $ 0.024 \icubang $ and $ 0.1 \icubang $.
The simulations are performed
using non-additive \abinitio
two-body, three-body, and four-body PESs for \parahyd
previously published by this research group.
\cite{ph2cluster:14faru,threebody:22ibra,fourbody:24ibra}
We show how the inclusion or omission of the different PESs
during the sampling and estimation phases of the PIMC simulation
affect the energy-density relationship,
the pressure-density relationship,
and the position distribution of molecules in the solid.
We also show how the separate energy components
(kinetic, two-body potential, three-body potential, and four-body potential)
change as a function of density.

The remainder of this paper is structured as follows:
In Sec.~(\ref{sec:model}),
we present the model of the solid,
including the Hamiltonian that describes it,
a description of the three \abinitio PESs used in the simulations,
and the sampling strategies used in the work.
In Sec.~(\ref{sec:computation}),
we briefly discuss the PIMC simulation technique,
and how we correct the systemic
Trotter and finite system size errors
inherent to performing PIMC simulations of solids.
In Sec.~(\ref{sec:results}),
we present and discuss the findings of our work.
Finally,
in Sec.~(\ref{sec:conclusion}),
we present both our conclusions as well as
suggestions for future work.

%%%%%%%%%%%%%%%%%%%%%%%%%%%%%%%%%%%%%%%%%%%%
%%% SECTION
%%%%%%%%%%%%%%%%%%%%%%%%%%%%%%%%%%%%%%%%%%%%

\section{Model} \label{sec:model}

%%%%%%%%%%%%%%%%%%%%%%%%%%%%%%%%%%%%%%%%%%%%
%%% HAMILTONIAN
%%%%%%%%%%%%%%%%%%%%%%%%%%%%%%%%%%%%%%%%%%%%

\subsection{Hamiltonian} \label{subsec:hamiltonian}

We model the solid
as a collection of $ N = 180 $ pointwise \para molecules in three spatial dimensions.
The molecules are placed at \hcp lattice sites
within a rectangular prism,
with periodic boundary conditions
applied in all three dimensions.
The Hamiltonian for the system is
\begin{align}
    \hat{H}
    &=
    - \frac{\hbar^2}{2 m}
    \sum_{i = 1}^{N}
    \nabla_i^2
    +
    \sum_{i < j}^{N} \hat{V}_2 (r_{ij})
    \nonumber \\
    &+
    \sum_{i < j < k}^{N} \hat{V}_3 (r_{ij}, r_{ik}, r_{jk})
    \nonumber \\
    &+
    \sum_{i < j < k < l}^{N} \hat{V}_4 (r_{ij}, r_{ik}, r_{il}, r_{jk}, r_{jl}, r_{kl})
\end{align}
\noindent
where
$ m $ is the mass of each \parahyd molecule,
$ r_{ij} $ is the relative distance between the centres of
mass of molecules with labels $ i $ and $ j $,
and $ \hat{V}_2 $, $ \hat{V}_3 $ and $ \hat{V}_4 $
are the two-body, three-body, and four-body PESs,
respectively.
\cite{ph2cluster:14faru,threebody:22ibra,fourbody:24ibra}

We model the \parahyd molecules as pointwise particles
interacting with one another through isotropic potentials.
The two-body PES implicitly accounts for rotational and vibrational
degrees of freedom in an adiabatic manner,
while the three-body and four-body PESs omit them entirely.
This is a reasonable assumption to make while
in phase I of the $ P \sh T $ phase region of $ H_2 $,
which,
at $ T = 4.2 \kel $,
extends from $ P = 0 \gpa $ up to around $ P = 60 \gpa $.
\cite{ph2solidexp:20greg}
Even at the highest densities explored in this work,
our simulations do not exceed pressures of $ 22 \, \mathrm{GPa} $.
In phases at higher pressures,
rotational degrees of freedom beyond
the adiabatic hindered rotor component become more important,
\cite{marx1999path,ph2solidexp:05gonc}
and the vibrationally averaged bond length decreases.
\cite{pathinteg:87cepe}
As a result,
we would need PESs with explicit rotational and vibrational degrees of freedom
to accurately simulate solid \parahyd to even higher pressures.

%%%%%%%%%%%%%%%%%%%%%%%%%%%%%%%%%%%%%%%%%%%%
%%% POTENTIAL ENERGY SURFACES
%%%%%%%%%%%%%%%%%%%%%%%%%%%%%%%%%%%%%%%%%%%%

\subsection{Potential Energy Surfaces} \label{subsec:pes}

The PIMC simulations in this work are performed with
three different \abinitio PESs for \parahyd.
The energies used in the training data
for each of these PESs
were collected from electronic structure calculations.
All the energies were calculated
at the correlated coupled-cluster theory level,
with singles, doubles, and perturbative triples excitations (CCSD(T)).
The calculations were performed
using augmented correlation-consistent polarized valnce zeta basis sets of cardinality $ N $ (AVNZ),
where $ N = \{D, T, Q, ...\} $.
Each calculation is supplemented by
a $ (3s3p2d) $ midbond basis
at the centre of mass of the atoms.

For the two-body interaction,
we use the Faruk-Schmidt-Hinde (FSH) potential.
This PES was created in 2014,
\cite{ph2cluster:14faru,ph2cluster:15schm}
when Faruk \etal
applied the adiabatic hindered rotor method
\cite{hinderedrotor:10li,hinderedrotor:10li_2,dhe:12wang,wang2013new,li2013analytic}
to a full six-dimensional \ce{H2}--\ce{H2} PES
published by Hinde.
\cite{h2pes:08hind}
This is an adiabatic treatment in which
the centre of mass pair distance is treated as the ``slow'' coordinate,
while the remaining vibrational and rotational degrees of freedom
are treated as the ``fast'' coordinates.
The result is an isotropic 1D PES
which treats the two \parahyd molecules in a pointwise manner,
and whose only parameter is the centre of mass distance between the two molecules.
In previous studies,
the FSH potential has successfully predicted the
experimentally observed first vibrational shifts of small \parahyd molecule clusters.
\cite{ph2cluster:14faru,ph2cluster:15schm,moleculeh2:20marr,ph2cluster:22schm}
The energies used in the original 6D \ce{H2}--\ce{H2} PES by Hinde
were calculated at the CCSD(T) level,
using AVQZ atom-centred basis sets.\cite{h2pes:08hind}
The two-body PES is implemented in the simulation using 1D linear interpolation.

For the three-body interaction,
we use a non-additive \abinitio three-body \parahyd interaction PES
published by this research group in 2022.
\cite{threebody:22ibra}
The energies used in the training data
were calculated at the CCSD(T) level,
using AVTZ atom-centred basis sets,
with the MRCC program (version 2019).\cite{elecstr:mrcc,elecstr:20kall}
The energies are isotropic,
and the PES is a function of only the three centre of mass pair distances
between the three \parahyd molecules.
The rigid rotor approximation removes
the three bond length degrees of freedom,
and the six-point Lebedev quadrature scheme
is used to spherically average over
the six rotational degrees of freedom.
The published three-body PES
uses a reproducing-kernel Hilbert space (RKHS) toolkit
to interpolate the training data.
\cite{rkhs:21unke}
For this work,
we change the implementation for the three-body PES
from the RKHS method to trilinear interpolation.
The training data for the new PES
is collected by evaluating the original PES
in a fine grid of Jacobi coordinates.
This change in implementation drastically improves
the evaluation time of the three-body interaction,
at the expense of increasing the required amount of memory.
Evaluating both three-body PESs
on PIMC simulation snapshots from three different densities
($ \rho = 0.024 \icubang $, $ 0.062 \icubang $, $ 0.1 \icubang $)
shows that the estimates of the average three-body potential energy per molecule
between the two implementations
are within $ 0.003 \wvn $, $ 0.01 \wvn $, and $ 0.08 \wvn $,
respectively.
The fit error is still much smaller than the systemic
finite basis set size error
from using the AVTZ basis set.
\cite{threebody:22ibra}

For the four-body interaction,
we use an \abinitio four-body \parahyd interaction PES
recently published by this research group.
\cite{fourbody:24ibra}
The energies in the training data
were calculated at the CCSD(T) level,
using AVDZ atom-centred basis sets,
with the MRCC program (version 2019).\cite{elecstr:mrcc,elecstr:20kall}
As is the case for the three-body PES,
the energies in the four-body PES are isotropic,
being a function of only the six centre of mass pair distances
between the four \parahyd molecules.
We use the rigid rotor appoximation and
the six-point Lebedev quadrature scheme
to remove the vibrational and rotational degrees of freedom.
The four-body PES is implemented using a feed-forward neural network in C++ libtorch.
\cite{pytorch:19pasz}

There are certain regions of coordinate space
for which the accuracy of the aforementioned PESs
are unknown or limited.
However,
for the range of conditions explored in these simulations,
the effect of these limitations on the results is minimal.
For example,
all three PESs were built using \abinitio training data
with side lengths greater than $ 2.2 \ang $,
and each of them relies on exponential short-range extrapolations
for samples with shorter side lengths.
However,
even at the highest explored density of $ 0.1 \icubang $,
our simulations spend very little time at such short distances.
At lower densities,
pair distances below $ 2.2 \ang $ are almost completely absent.
As another example,
consider a collection of four \parahyd molecules where
some of the side lengths are
very small (around $ 2.2 \ang $ or shorter)
and others are
much larger (around $ 4.5 \ang $ or longer).
For such a cluster,
the four-body PES
produces an interaction energy that is a physically unrealistic
linear combination of short-range and long-range interaction energies.
\cite{fourbody:24ibra}
However,
under these conditions
this four-body energy is also typically
four to five orders of magnitude smaller than
the total two-body and three-body energies of this same cluster,
and thus we can easily neglect the inaccuracy.

%%%%%%%%%%%%%%%%%%%%%%%%%%%%%%%%%%%%%%%%%%%%
%%% SAMPLING STRATEGIES
%%%%%%%%%%%%%%%%%%%%%%%%%%%%%%%%%%%%%%%%%%%%

\subsection{Sampling strategies} \label{subsec:sampling}

During a simulation,
the four-body PES takes longer to evaluate than the three-body PES,
which itself takes longer to evaluate than the two-body PES.
One reason for this increase in runtime is the combinatoric increase
in the number of required terms.
When we evaluate a $ k $-body PES on a system of $ N $ particles,
the number of groups of $ k $ particles grows as $ \mathcal{O}(N^k) $.
Another reason the evaluations take longer is that
the implementations for the higher-order many-body PESs become more expensive.
The two-body PES is implemented using a very fast 1D linear interpolation.
The three-body PES is implemented using trilinear interpolation,
which requires several times as many calculations as 1D linear interpolation.
The four-body PES is implemented using a feed-forward neural network,
which is implemented as a repeated sequence of matrix multiplications
and activation function calls.
The inclusion of the higher-order many-body interaction PESs
is expensive enough to the point that simulations under certain conditions
become intractable.

We can use approximation sampling strategies
to improve the runtime performance of our simulations.
Each of the PESs can be used in one of two ways during the simulation;
sampling and estimating.
During sampling,
we use the PESs to calculate the energies
for the current and proposed positions
for a Monte Carlo step,
and use these energies to determine if we accept of reject the proposed position.
During estimating,
we use the PESs to calculate the total interaction energy of the system,
which we use as a sample
for the estimator of the total interaction energy.
As an approximation,
we can choose to omit certain higher-order many-body PESs
from the sampling stages of the simulation,
but keep them in the estimating stages of the simulation.
This approximation assumes that the higher-order many-body interactions
have a negligible or perturbative effect on the molecular positions
of the \parahyd molecules during sampling.
In a study on the effect of three-body interactions in solid helium
by Barnes and Hinde,
\cite{pathinteg:17barn_1}
and a study on the effect of three-body interactions in solid \parahyd
by this research group,
\cite{pathinteg:22ibra}
it was shown that this assumption
is valid except at very high densities.

With our three PESs,
we can define three sampling strategies,
each one defined by which of the higher-order many-body PESs
are omitted from the sampling phase of the simulation.
In the first sampling strategy,
labelled by \sampT,
we use only the two-body PES for the sampling steps.
In other words,
the potential energy component of the change in energy $ \Delta V $
used to determine if a proposed Monte Carlo step is accepted or rejected
is given by
\begin{equation} \label{eq:sampling:sampT}
    \Delta V
    =
    V_2(\vb{q}_f)
    -
    V_2(\vb{q}_i) \, ,
\end{equation}
\noindent
where
$ V_2 $ is the two-body interaction potential,
and $ \vb{q}_f $ and $ \vb{q}_i $
are the final and initial positions of all molecules in the solid,
respectively.
In the next sampling strategy,
labelled by \sampTT,
we use only the two-body and three-body PESs for the sampling steps.
The potential energy difference during sampling for the \sampTT strategy is
\begin{equation} \label{eq:sampling:sampTT}
    \Delta V
    =
    \left[ V_2(\vb{q}_f) + V_3(\vb{q}_f) \right]
    -
    \left[ V_2(\vb{q}_i) + V_3(\vb{q}_i) \right] \, .
\end{equation}
\noindent
In the last sampling strategy,
labelled by \sampTTF,
we use the two-body, three-body, and four-body PESs for the sampling steps.
The potential energy difference during sampling for the \sampTTF strategy is
\begin{align} \label{eq:sampling:sampTTF}
    \Delta V
    &=
    \left[ V_2(\vb{q}_f) + V_3(\vb{q}_f) + V_4(\vb{q}_f) \right]
    \nonumber \\
    &-                                     
    \left[ V_2(\vb{q}_i) + V_3(\vb{q}_i) + V_4(\vb{q}_i) \right] \, .
\end{align}
\noindent
The estimation and sampling stages of the simulations are separate from one another.
Thus,
in each of the \sampT, \sampTT, and \sampTTF stratgies,
we can still use all three PESs during the estimation stage.
For example,
in the \sampT simulations,
we only use the two-body PES to calculate the energies for
the current and proposed positions during the Monte Carlo steps,
but we still use the
two-body, three-body, and four-body PESs
to calculate the total interaction energies
during the estimation stage.

The four-body PES is much more expensive to evaluate compared to
the two-body and three-body PESs.
For the \sampT and \sampTT cases,
we are able to reasonably estimate the average four-body potential energy
by calling its estimator on only 40 to 50 state snapshots.
This is because
the variance of the four-body potential energy estimator is relatively small
for all densities explored.
However,
its expensive evaluation time
makes the \sampTTF simulations very slow,
and thus we decide to perform them
only at the relatively high density of $ 0.1 \icubang $.
To perform the \sampTTF simulations,
we take 980 decorrelated position snapshots of the system from a \sampT simulation,
and use them as initial positions for 980 separate \sampTTF simulations.
At this density,
the \sampT and \sampTTF simulations
have similar acceptance rates for the same Monte Carlo step sizes.
To err on the side of caution,
each \sampTTF simulation is run for twice the autocorrelation time
of the \sampT simulation before calling the estimators,
and only a single estimate is performed before a simulation ends.

We should note that this choice of density was made
after having already performed the \sampT and \sampTT simulations,
with full knowledge that the pressure for the \sampT and \sampTT simulation results
already exceed the experimental results at this density (see Sec.~(\ref{subsec:pressure_density})).
The emphasis is on choosing a density
where we expect the effect of the four-body PES on the solid structure to be very strong,
and the change in the radial distribution function (discussed in Sec.~(\ref{subsec:radial_pair_distribution}))
to be noticeable.

%%%%%%%%%%%%%%%%%%%%%%%%%%%%%%%%%%%%%%%%%%%%
%%% COMPUTATION
%%%%%%%%%%%%%%%%%%%%%%%%%%%%%%%%%%%%%%%%%%%%

\section{Computation} \label{sec:computation}

%%%%%%%%%%%%%%%%%%%%%%%%%%%%%%%%%%%%%%%%%%%%
%%% PATH-INTEGRAL MONTE CARLO
%%%%%%%%%%%%%%%%%%%%%%%%%%%%%%%%%%%%%%%%%%%%

\subsection{Path-integral Monte Carlo} \label{subsec:pimc}

Suppose we have a system described by a Hamiltonian $ \hat{H} $
at a finite temperature $ \beta = 1 / k_B T $.
The expectation value of a quantum mechanical operator $ \hat{A} $
in this system can be expressed using the PIMC method as
\begin{align} \label{eq:pimc:expectation_value}
    \left\langle
        \hat{A}
    \right\rangle
    &=
    \frac{1}{Z}
    Tr
    \left\{
        \hat{A} e^{- \beta \hat{H}}
    \right\}
    \nonumber \\
    &=
    \frac{1}{Z}
    \int \dd{\vb{q}} \dd{\vb{q}'}
    \mel{ \vb{q}          }{ \hat{A}             }{ \vb{q}^{\prime} }
    \mel{ \vb{q}^{\prime} }{ e^{- \beta \hat{H}} }{ \vb{q}          }
    \, ,
\end{align}
\noindent
where
$ Z = Tr \left\{ \hat{A} e^{- \beta \hat{H}} \right\} $ is the partition function
and $ \vb{q} $ represents the positions of all particles in the system.
We can discretize the integral in imaginary time,
writing
\begin{equation} \label{eq:pimc:discretize}
    \mel{ \vb{q}^{\prime} }{ e^{ - \beta \hat{H} } }{ \vb{q} }
    =
    \int
    \prod_{i=2}^{P} \dd \vb{q}_i \times \prod_{i=1}^{P}
    \mel{ \vb{q}_{i} }{ e^{ - \tau \hat{H} } }{ \vb{q}_{i+1} } ,
\end{equation}
\noindent
where $ \tau = \beta / P $ is the imaginary time step,
and $ \vb{q}_i $ represents
the positions of all particles in the system at the imaginary time step with label $ i $.
We also apply the boundary condition $ \vb{q}_1 = \vb{q}^{\prime} $ and $ \vb{q}_{P+1} = \vb{q} $
to Eq.~(\ref{eq:pimc:discretize}).
The integral
thus forming a ring of coordinates $ \{ \vb{q}_i \} $.
We use the PIMC method
to sample the integral Eq.~(\ref{eq:pimc:expectation_value}).
\cite{pathinteg:95cepe}

Because \parahyd is a boson,
our simulation method should in principle account for exchange effects,
using a strategy such as the worm algorithm.
\cite{pathinteg:06bonia,pathinteg:06bonib}
However,
previous simulation studies
indicate that bosonic exchange between \parahyd molecules
is almost completely suppressed in the solid phase.
\cite{h2pes:13omiy,ph2solidtheo:17duss,pathinteg:17boni,pathinteg:19ibra}
Thus in practice,
the \parahyd molecules can be treated as distinguishable particles.

%%%%%%%%%%%%%%%%%%%%%%%%%%%%%%%%%%%%%%%%%%%%
%%% TROTTER ERROR
%%%%%%%%%%%%%%%%%%%%%%%%%%%%%%%%%%%%%%%%%%%%

\subsection{Trotter error} \label{subsec:trotter_error}

The PIMC method involves separating the Hamiltonian
in Eq.~(\ref{eq:pimc:discretize})
into its potential and kinetic components.
Upon doing so,
we introduce a systematic Trotter factorization error into our expectation values.
\cite{pathinteg:95cepe,pathinteg:17yan}
This systemic error grows with the imaginary time step $ \tau $.
In the limit $ \tau \rightarrow 0 $,
or in other words,
in the limit of infinitely many time steps,
this systemic error vanishes.
We account for the Trotter factorization error
in the same way that we did in a previous study
on the effects of three-body interactions in solid \parahyd.
\cite{pathinteg:22ibra}

One method to eliminate the factorization error is the $ \tau $-extrapolation method.
We perform several simulations under the same conditions
(the same temperature, density, number of particles, \textit{etc.}),
with the only difference between these simulations being the number of time slices $ P $.
Each simulation
is run with a different value of $ \tau $,
and produces a different expectation value $ \langle \hat{A} \rangle (\tau) $.
We can fit these expectation values using
\begin{equation} \label{eq:trotter_error:fit_extrapolation}
    \langle \hat{A} \rangle (\tau)
    =
    \langle \hat{A} \rangle_{0}
    +
    c_2 \tau^2
    +
    c_4 \tau^4
\end{equation}
\noindent
where
$ \langle \hat{A} \rangle_{0} $ is the expectation value of the observable $ \hat{A} $
in the limit of $ \tau \rightarrow 0 $,
and $ c_2 $ and $ c_4 $ are fit parameters.
\cite{pathinteg:17yan}
The $ \tau $-extrapolation method is ideal
when the kinetic and potential energy components are small in magnitude.
Under these conditions,
the Trotter factorization error is small,
and the extrapolation Eq.~(\ref{eq:trotter_error:fit_extrapolation})
is valid even for fairly large values of $ \tau $.
However,
as we increase the density of the solid,
the kinetic and potential energy components increase in magnitude,
and the fit in Eq.~(\ref{eq:trotter_error:fit_extrapolation})
becomes less suitable.

Another strategy to mitigate the factorization error is the small-$ \tau $ method,
where we select a value of $ \tau $ that is small enough
such that the Trotter factorization error is tolerably small.
This strategy is simpler to implement,
but leaves part of the systemic Trotter error in the result.

We use the $ \tau $-extrapolation method to calculate the energies
in a fine grid of densities near the equilibrium density of $ \rho_0 = 0.0261 \icubang $,
\cite{ph2solidtheo:17duss,ph2solidexp:23pris}
where the kinetic and potential energy components are relatively small.
For each density,
we perform simulations using $ P = \{ 64, 80, 96, 128, 192 \} $ time slices,
collect the expectation values for the energies,
and use Eq.~(\ref{eq:trotter_error:fit_extrapolation})
to find the expectation values in the $ \tau \rightarrow 0 $ limit.
This method allows us to recover low-variance energies
and accurately find the equilibrium density.
In the range of densities where we use this $ \tau $-extrapolation method,
the total energies span a range of about $ 2 \wvn $,
whereas the standard error of the mean of the resulting energies is only on the order of $ 0.01 \wvn $.

We use the small-$ \tau $ method to calculate the energies of the solid
in a sparse grid of densities between $ \rho = 0.024 \icubang $ and $ \rho = 0.1 \icubang $.
We choose $ P = 960 $ beads ($ \tau \approx 2.48 \times 10^{-4} \kel^{-1}$).
In an earlier similar study,
\cite{pathinteg:22ibra}
it was found that compared to the extrapolated value,
this choice of time step leads to a Trotter error in the
total energy of about
$ 0.2 \% $ at $ \rho = 0.026 \icubang $ and $ \rho = 0.04 \icubang $,
and about $ 2.5 \% $ at $ \rho = 0.1 \icubang $.
Also,
at very high densities the simulation results deviate from the experimental data
to such an extent that it dwarfs the systemic Trotter error.

%%%%%%%%%%%%%%%%%%%%%%%%%%%%%%%%%%%%%%%%%%%%
%%% FINITE SYSTEM SIZE ERROR
%%%%%%%%%%%%%%%%%%%%%%%%%%%%%%%%%%%%%%%%%%%%

\subsection{Finite system size error} \label{subsec:system_size_error}

Our simulations of solid \parahyd
are done with a finite sized box
with periodic boundary conditions applied to each dimension.
We must apply the minimum image rules
to make sure that we calculate the correct distance between
any two molecules.
For the two-body PES,
we apply the standard minimum image convention.
Interactions between any two molecules greater than
a cutoff distance of $ L / 2 $ apart are ignored,
where $ L $ is the shortest of the three side lengths of the simulation box.
For the three-body PES,
we follow the three-body minimum image convention described by Attard.
\cite{threebody:91atta}
For the four-body PES,
we adapt Attard's minimum image convention
by ignoring the interaction energy
of any four-body geometry where any of the side lengths is greater than $ L/2 $.

Because we perform our simulations within a finite sized simulation box,
the potential energy estimators are unable to account for interactions
involving molecules outside of this box.
As a result,
all the potential energy estimates
suffer from a systemic finite system size error.
To correct this error we use long-range tail corrections,
which are calculations that approximate the interaction energy
between a molecule inside of the simulation box
and all other molecules outside the simulation box.

The contribution to the two-body interaction energy per molecule
beyond a cutoff distance $ L / 2 $ is given by
\begin{equation} \label{eq:system_size_error:two_body_tail}
    \epsilon_2^{(t)}
    =
    4 \pi
    \frac{\rho}{2}
    \int_{L/2}^{\infty}
    \dd{r} \,
    r^2
    V_2(r)
    g(r)
\end{equation}
\noindent
where
$ g(r) $ is the radial pair distribution.
The simulation is only able to calculate $ g(r) $ within the cutoff distance.
In the pair tail correction,
we approximate the radial pair distribution outside the simulation box as being uniform,
and set $ g(r) = 1 $ in Eq.~(\ref{eq:system_size_error:two_body_tail}).

We calculate the three-body tail correction using
Eq.~(2.150) in Ref.~[\onlinecite{book:89alle}],
which accounts for interactions from all three-body geometries
where at least one side length is greater than $ L / 2 $.
The calculation requires the three-body distribution function,
which we approximate using a product of three radial pair distribution functions
collected from the simulation.

In an earlier work,
\cite{threebody:22ibra}
we performed \sampT PIMC simulations of solid \parahyd
using simulation boxes with both $ N = 180 $ particles and $ N = 448 $ particles,
with the two-body and three-body tail corrections applied to both.
The difference between the two total energies,
relative to the overall total energy,
was less than $ 0.2 \% $ even at the highest densities explored.
This was used to verify that $ N = 180 $ is a sufficiently large enough
number of particles to simulate the solid
insofar as energy estimates are concerned.

We do not apply a tail correction for the four-body interaction energy,
as we expect it to be negligible.
Even at the highest densities considered in this work,
where the tail corrections are at their most significant,
the two-body and three-body tail corrections
for a simulation box of $ N = 180 $ particles
are less than
$ 1.5 \% $ and $ 0.5 \% $ of the total interaction energy,
respectively.
\cite{threebody:22ibra}
We know that the magnitude of a tail correction depends on
the long-range interaction strength of the corresponding PES.
At long distances,
the four-body PES decays with $ r^{-12} $,
much more quickly than the two-body PES at $ r^{-6} $
or the three-body PES at $ r^{-9} $.
We can thus expect the four-body tail correction to be even smaller than the
three-body tail correction,
which itself is already very small.

%%%%%%%%%%%%%%%%%%%%%%%%%%%%%%%%%%%%%%%%%%%%
%%% RESULTS
%%%%%%%%%%%%%%%%%%%%%%%%%%%%%%%%%%%%%%%%%%%%

\section{Results} \label{sec:results}

%%%%%%%%%%%%%%%%%%%%%%%%%%%%%%%%%%%%%%%%%%%%
%%% ENERGY COMPONENTS
%%%%%%%%%%%%%%%%%%%%%%%%%%%%%%%%%%%%%%%%%%%%

\subsection{Energy components} \label{subsec:energy_components}

\begin{figure} [h]
    \centering
    \includegraphics[width=\linewidth]{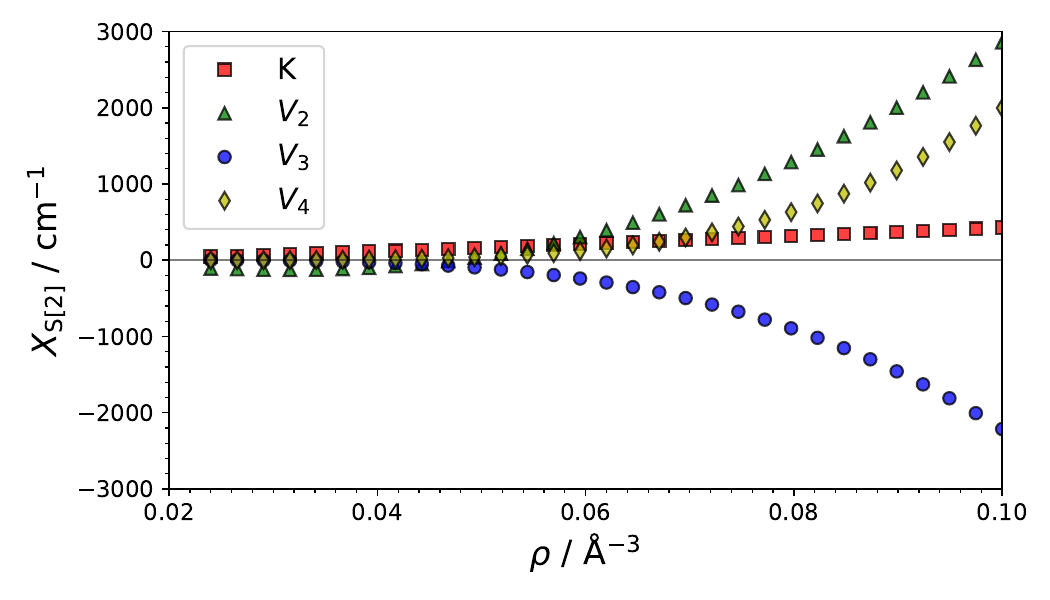}
    \caption{
        The energy per particle as a function of the density $ \rho $,
        separated into the
        kinetic (red, squares),
        two-body potential (green, triangles),
        three-body potential (blue, circles), and
        four-body potential (yellow, diamonds) energy components.
        The simulations are performed using the \sampT sampling strategy.
        The variable $ X_{\rm S[2]} $ on the vertical axis label
        is a placeholder for the various energy components.
        The standard error of the mean of the data is not visible
        on the scale of the figure.
    }
    \label{fig:energy_component_pert2b_coarse}
\end{figure}

In Fig.~(\ref{fig:energy_component_pert2b_coarse}),
we show how each of the four energy components varies as a function of density
in simulations performed using the \sampT strategy.
The two-body, three-body, and four-body potential energies all vary exponentially
at higher densities,
while the kinetic energy increases linearly.
Earlier studies on solid \parahyd
suggest that this linear trend for the kinetic energy only holds at lower densities.
\cite{h2pes:13omiy}

\begin{figure} [h]
    \centering
    \includegraphics[width=\linewidth]{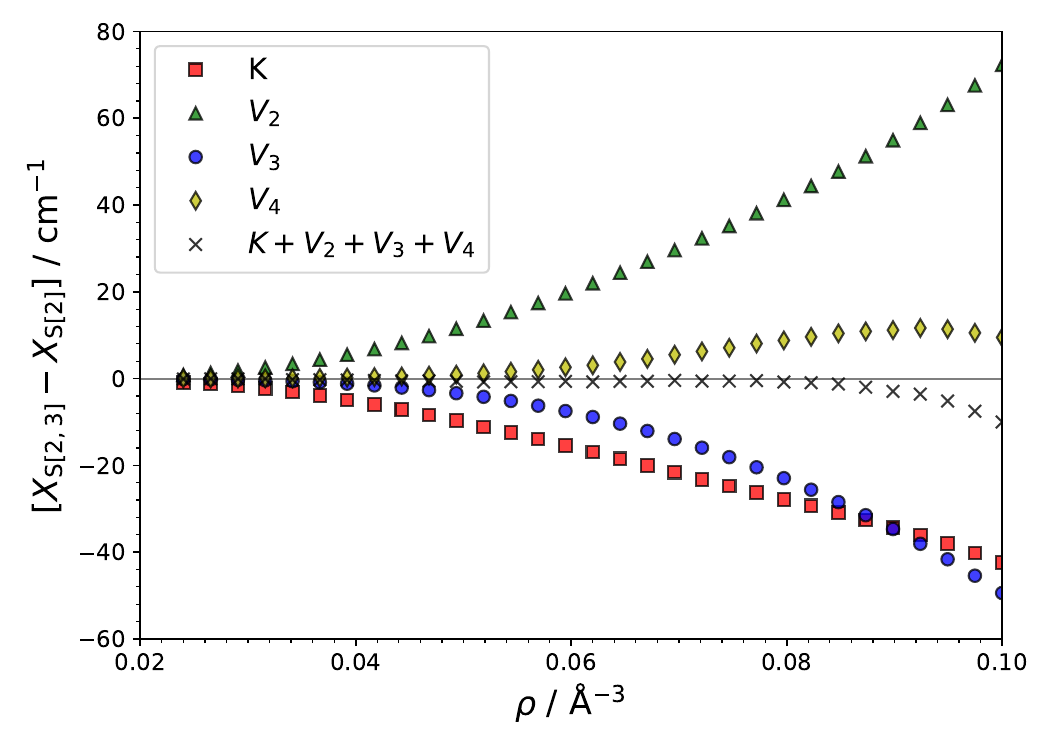}
    \caption{
        The difference in the energy per particle as a function of the density $ \rho $,
        as collected from the \sampT and \sampTT simulations.
        In addition to the separate
        kinetic (red, squares),
        two-body potential (green, triangles),
        three-body potential (blue, circles), and
        four-body potential (yellow, diamonds) energy components,
        we also show the difference between the total energy (black, crosses).
        The variables $ X_{\rm S[2]} $ and $ X_{\rm S[2,3]} $
        on the vertical axis label
        are placeholders for the energy components
        collected from \sampT and \sampTT simulations,
        respectively.
        The standard error of the mean of the data is not visible
        on the scale of the figure.
    }
    \label{fig:energy_component_differences_pert2b3b_pert2b_coarse}
\end{figure}

The inclusion or omission of higher-order many-body interactions during sampling
affects the position distribution of the molecules in the solid,
which in turn
affects estimates of each of the four energy components.
To demonstrate this,
we perform PIMC simulations of solid \parahyd using both
the \sampT and \sampTT sampling strategies,
for densities between $ 0.024 \icubang $ and $ 0.1 \icubang $.
We then take the individual energy components from the \sampTT simulations,
subtract their counterparts from the \sampT simulations,
and plot the differences as a function of density in Fig.~(\ref{fig:energy_component_differences_pert2b3b_pert2b_coarse}).

Each of the four energy components experiences a shift
when switching from the \sampT strategy to the \sampTT strategy.
The kinetic and three-body potential energies decrease,
while the two-body potential and four-body potential energies increase.
For the four-body interaction energy,
the trend of the difference increasing reverses at the highest densities explored here.
We are unsure of why this is the case.
The total energy is almost completely unchanged
except near $ 0.1 \icubang $,
where the difference is just under $ 0.3 \% $ of the total interaction energy.
The changes in the individual energy components almost completely cancel each other out.

We can perform a $ \tau $-extrapolation of the kinetic energy component on its own
at $ \rho_0 = 0.0261 \icubang $.
The predicted average kinetic energy per molecule at the equilibrium density at $ T = 4.2 \kel $ is
$ \ev{K} = 53.94(1) \wvn $ for the \sampT simulations, and
$ \ev{K} = 52.79(1) \wvn $ for the \sampTT simulations.
Both prediction are greater than the experimental result of $ 49.3(8) \wvn $
at $ T = 5 \kel $.
\cite{ph2solidexp:23pris}

In Table.~(\ref{tab:energies_at_highest_density}),
we show the estimates of the energy components collected from
the \sampT, \sampTT, and \sampTTF simulations
performed using $ P = 960 $ beads
at a density of $ 0.1 \icubang $.
As is the case in the \sampTT simulations,
the \sampTTF simulations see
a decrease in the kinetic and three-body potential energies,
and an increase in the two-body potential energy.
However,
the shifts are even greater in the \sampTTF simulations
than in the \sampTT simulations.
The four-body potential energy is very similar in all three sampling cases.
The total energy is very similar in all three cases as well,
only being about $ 1 \% $ larger in the \sampTTF case compared to the \sampT case.

\begin{table}
    \caption{
        In order from top to bottom,
        estimates of the average
        kinetic,
        two-body potential,
        three-body potential, and
        four-body potential energy per molecule,
        followed by the sum of the previous four values.
        The results are shown from simulations performed
        under the \sampT, \sampTT, and \sampTTF sampling strategies,
        performed at $ 0.1 \icubang $.
        All energies are shown in units of $ \wvn $.
        The two-body and three-body interaction energies
        have had their respective long-range tail corrections applied.
    }
    \begin{ruledtabular}
        \begin{tabular}{lccc}
                & \sampT             & \sampTT            & \sampTTF           \\
        \hline
        $ K   $ & $   426.969(34)  $ & $   384.562(41)  $ & $   361.47(25)   $ \\
        $ V_2 $ & $  2856.5974(81) $ & $  2928.937(13)  $ & $  3058.842(96)  $ \\
        $ V_3 $ & $ -2215.1549(35) $ & $ -2264.5788(68) $ & $ -2324.996(47)  $ \\
        $ V_4 $ & $  1996.40(11)   $ & $  2005.88(14)   $ & $  1999.118(37)  $ \\
        total   & $  3064.81(13)   $ & $  3054.80(15)   $ & $  3094.44(27)   $ \\
        \end{tabular}
    \end{ruledtabular}
    \label{tab:energies_at_highest_density}
\end{table}

%%%%%%%%%%%%%%%%%%%%%%%%%%%%%%%%%%%%%%%%%%%%
%%% ENERGY-DENSITY EQUATION OF STATE
%%%%%%%%%%%%%%%%%%%%%%%%%%%%%%%%%%%%%%%%%%%%

\subsection{Energy-density equation of state} \label{subsec:energy_density}

We perform PIMC simulations of solid \parahyd at $ T = 4.2 \kel $
using the \sampT and \sampTT sampling strategies.
One set of simulations is performed in a fine,
regularly spaced grid of densities between
$ 0.025 \icubang $ and $ 0.027 \icubang $,
around the equlibrium density region.
For these simulations,
we handle the Trotter error using the $ \tau $-extrapolation method.
Another set of simulations is performed in a coarser,
regularly spaced grid of densities between
$ 0.024 \icubang $ and $ 0.1 \icubang $,
for which we handle the Trotter error using the small-$ \tau $ method.

During the estimation phase,
we use the primitive estimator for the kinetic energy,
and the standard estimators for the two-body, three-body and four-body
interaction potential energies,
taking periodic boundary conditions into account.
From the estimates,
we can create three energy-density curves.
One energy-density curve is the sum of only the kinetic and two-body interaction energies,
which we label with \estiT.
Another is the sum of the kinetic, two-body, and three-body energies,
which we label with \estiTT.
The final is the sum of the kinetic, two-body, three-body, and four-body energies,
which we label with \estiTTF.

We can fit the average energy per particle as a function of density
for the \estiT, \estiTT, and \estiTTF cases
to a modified Birch equation of state,\cite{ph2solidexp:79drie,ph2solidexp:00cohe}
\begin{equation} \label{eq:energy_density:modified_birch_eos}
    \epsilon
    =
    \epsilon_0
    -
    \frac{P_0}{d \rho_0}
    +
    \frac{1}{\rho_0}
    \sum_{n=1}^{4}
    \kappa_{n} d^{2n / 3} \, ,
\end{equation}
\noindent
where $ d = \rho / \rho_0 $ is the number density normalized
by the experimental value of the equilibrium density $ \rho_0 = 0.0261 \icubang $.
This normalization is done to reduce the disparity in the orders of magnitude of the fit parameters.
In the supplementary material,
we show the fit parameters
for the \estiT, \estiTT, and \estiTTF cases,
calculated using both the \sampT and \sampTT strategies.

\begin{figure} [h]
    \centering
    \includegraphics[width=\linewidth]{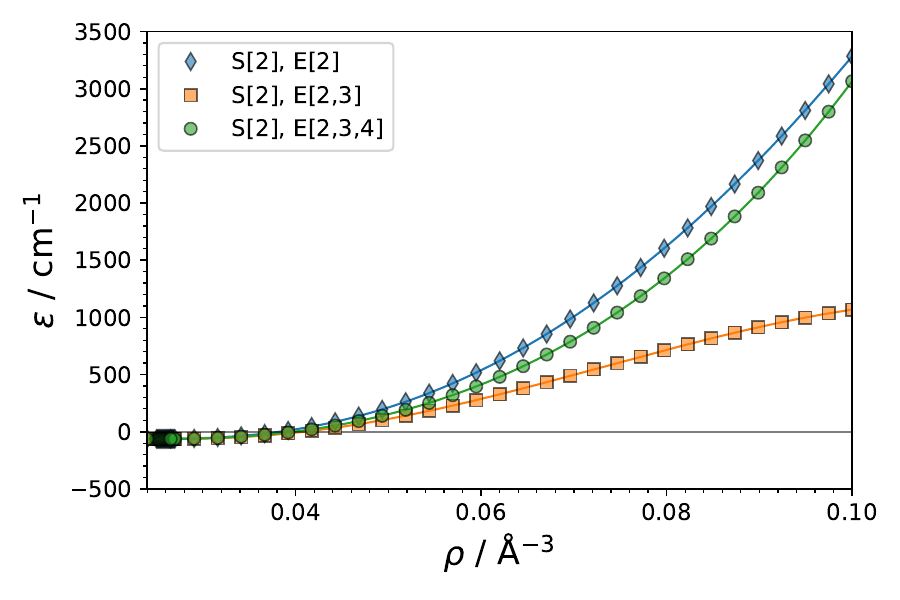}
    \caption{
        The energy per molecule $ \epsilon $ as a function of the density $ \rho $
        of solid parahydrogen
        for the
        \estiT (blue, diamonds),
        \estiTT (orange, squares),
        and \estiTTF (green, circles) estimates.
        All results are shown for simulations performed with the \sampT sampling strategy.
        The curves are found by fitting the results
        to Eq.~(\ref{eq:energy_density:modified_birch_eos}).
        The standard error of the mean of the data is not visible
        on the scale of the figure.
    }
    \label{fig:energy_vs_density_eos_coarse_pert2b}
\end{figure}

In Fig.~(\ref{fig:energy_vs_density_eos_coarse_pert2b}),
we plot the energy-density results collected from simulations
performed with the \sampT sampling strategy.
These energy-density curves are nearly indistinguishable from those collected from the \sampTT simulations.
The energy-density curves for the
\estiT, \estiTT, and \estiTTF cases
follow the expected trends based on their interaction strengths at short intermolecular separations.
For the \estiT curve,
the total energy increases as a function of density.
The three-body interaction is exponentially attractive at short distances,
and when we include it to form the \estiTT curve,
the total energy decreases.
Similarly,
the four-body interaction is exponentially repulsive at short distances,
and including it to form the \estiTTF curve causes the
total energy to once again increase.

\begin{figure} [h]
    \centering
    \includegraphics[width=\linewidth]{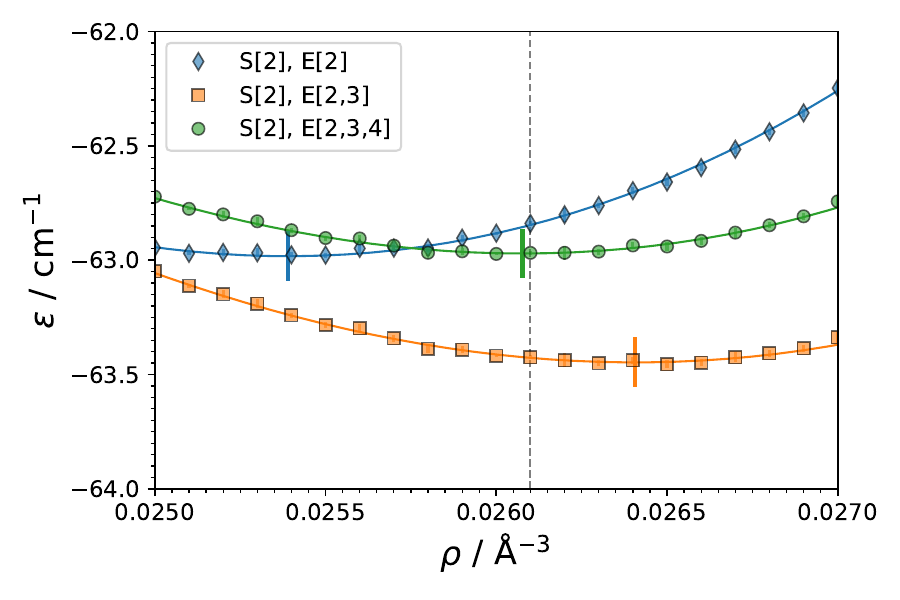}
    \caption{
        The energy per molecule $ \epsilon $ as a function of the density $ \rho $
        of solid parahydrogen around the equilibrium density
        for the
        \estiT (blue, diamonds),
        \estiTT (orange, squares),
        and \estiTTF (green, circles) estimates.
        All results are shown for simulations performed with the \sampT sampling strategy.
        The short vertical
        blue, orange, and green lines
        indicate the equilibrium density predicted by the
        \estiT, \estiTT, and \estiTTF estimates, respectively.
        The vertical dashed line is located at $ \rho_0 = 0.0261 \icubang $,
        the experimental result of the equlibrium density.
        \cite{ph2solidtheo:17duss,ph2solidexp:23pris}
        The statistical uncertainties are visible within the
        semitransparent faces of the markers.
    }
    \label{fig:energy_vs_density_eos_equilibrium_pert2b}
\end{figure}

\begin{figure} [h]
    \centering
    \includegraphics[width=\linewidth]{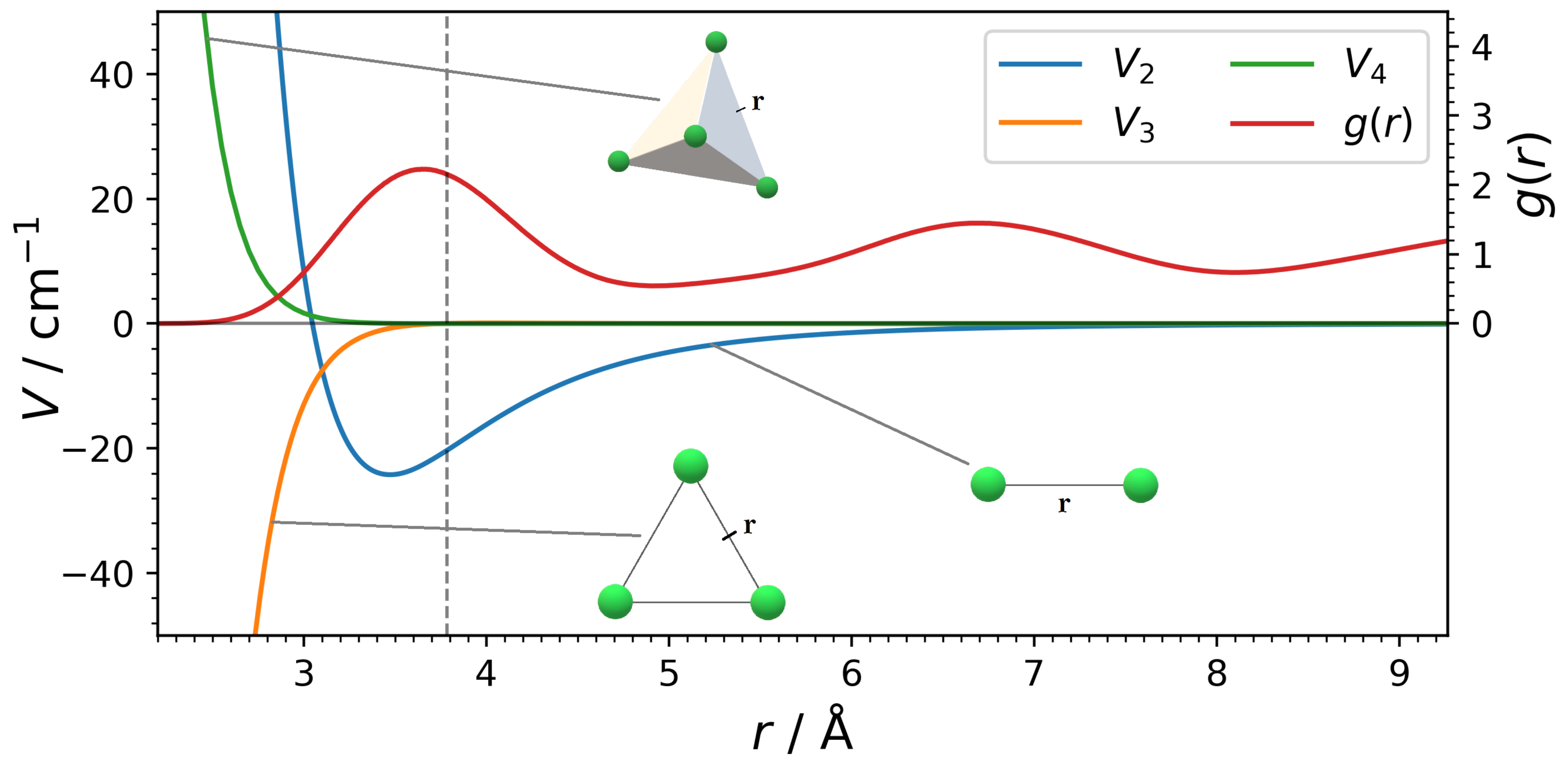}
    \caption{
        The radial pair distribution function $ g(r) $ (red)
        of solid \parahyd at $ \rho_0 = 0.0261 \icubang $,
        collected from an \sampT simulation.
        Also shown are
        the interaction energies for particular geometries in the \textit{hcp} lattice.
        We show
        the two-body potential energy
        for two \parahyd molecules a distance $ r $ apart (blue),
        the non-additive three-body potential energy
        for three \parahyd molecules at the corners of an equilateral triangle of side length $ r $ (orange),
        and
        the non-additive four-body potential energy
        for four \parahyd molecules at the corners of a perfect tetrahedron of side length $ r $ (green).
        The vertical dashed line shows the position of the lattice constant at this density.
    }
    \label{fig:relevant_distances}
\end{figure}

In Fig.~(\ref{fig:energy_vs_density_eos_equilibrium_pert2b}),
we show the energy-density curves
from the \sampT simulations
around the equilibrium density region.
The \estiT and \estiTT curves
are shifted downwards compared to their counterparts
presented in Ref.~[\onlinecite{pathinteg:22ibra}],
due to an error in the calculation of the pair tail correction in the earlier work.
Even around the equilibrium density,
where the lattice constant is about $ 3.8 \ang $,
the three-body and four-body interactions
have a small but measurable effect
on the total interaction energy.
As demonstrated in Fig.~(\ref{fig:relevant_distances})
for specific geometric configurations,
this behaviour would be unreasonable from a classical model of solid \parahyd.
At this lattice constant
the three-body and four-body interactions are negligibly weak
compared to the two-body interaction.
However,
the large zero-point quantum displacement of the \parahyd molecules
allows their probability distributions to reach distances
where these many-body interactions are significant.
The predicted equilibrium densities from the \sampT simulations are
$ 0.02539 \icubang $ for the \estiT case,
$ 0.02640 \icubang $ for the \estiTT case, and
$ 0.02608 \icubang $ for the \estiTTF case.
The inclusion of the four-body interaction in the total energy estimator
improves the prediction of the equilibrium density.
The prediction given by the \estiTTF is closer to the experimental value
than either the \estiT or \estiTT curves.

When limited to the
energy and density scales relevant near the equilibrium density,
the differences in the total energies from the
\sampT and \sampTT simulations become noticeable,
but are still very small.
We show the energy-density curves alongside one another
in the supplementary material.
The predicted equilibrium densities for each of the \estiT, \estiTT, and \estiTTF estimations
are essentially unchanged in the \sampTT simulations.
Table~(\ref{tab:energy_per_molecule_at_equilibrium})
shows the average energy per molecule at the equilibrium density
for each sampling strategy and estimation case.
These are in line with estimates from
PIMC simulations performed using
the Moraldi, Buck, and Silvera-Goldman potentials.
\cite{h2pes:13omiy}

\begin{table} [h]
    \begin{ruledtabular}
    \begin{tabular}{lcc}
             & \sampT        & \sampTT       \\
    \hline
    \estiT   & $ -62.98(1) $ & $ -62.91(2) $ \\
    \estiTT  & $ -63.45(1) $ & $ -63.53(2) $ \\
    \estiTTF & $ -62.97(1) $ & $ -63.04(2) $ 
    \end{tabular}
    \end{ruledtabular}
    \caption{
        The average energy per molecule at the respective equilibrium densities (provided in the text)
        of each energy-density curve.
        Values are in units of $ \wvn $.
    }
    \label{tab:energy_per_molecule_at_equilibrium}
\end{table}

%%%%%%%%%%%%%%%%%%%%%%%%%%%%%%%%%%%%%%%%%%%%
%%% PRESSURE-DENSITY EQUATION OF STATE
%%%%%%%%%%%%%%%%%%%%%%%%%%%%%%%%%%%%%%%%%%%%

\subsection{Pressure-density equation of state} \label{subsec:pressure_density}

We can calculate the pressure as a function of density of solid \parahyd
from the energy-density fit Eq.~(\ref{eq:energy_density:modified_birch_eos}) using
\begin{equation} \label{eq:pressure_density:pressure_modified_birch_eos}
    P
    =
    \rho^2
    \eval{\pdv{\epsilon}{\rho}}_{T}
    =
    P_0 + \frac{2}{3} \sum_{n=1}^{4} n \kappa_n d^{(2n + 3) / 3} \, .
\end{equation}

\begin{figure} [h]
    \centering
    \includegraphics[width=\linewidth]{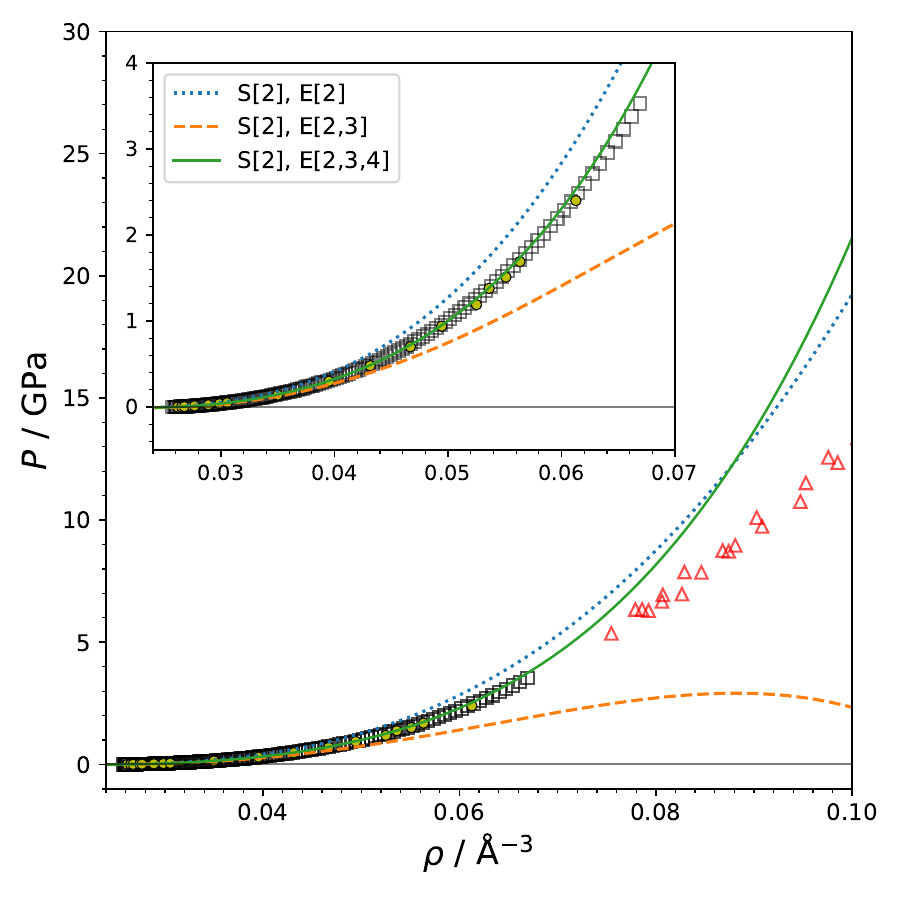}
    \caption{The pressure $ P $ as a function of the density $ \rho $
        of solid parahydrogen
        for the
        \estiT (blue, dotted line),
        \estiTT (orange, dashed line), and
        \estiTTF (green, solid line) estimates.
        The curves are calculated using
        Eq.~(\ref{eq:pressure_density:pressure_modified_birch_eos}).
        Also shown are experimentally derived results from
        Ref.~[\onlinecite{ph2solidexp:83ishm}] (solid yellow, circles),
        Refs.~[\onlinecite{ph2solidexp:80silv},\onlinecite{ph2solidexp:79drie}] (black, squares), and
        Ref.~[\onlinecite{ph2solidexp:94mao}] (red, triangles).
    }
    \label{fig:pressure_vs_density_eos_coarse_pert2b}
\end{figure}

In Fig.~(\ref{fig:pressure_vs_density_eos_coarse_pert2b}),
we show the pressure-density curves for simulations performed using the \sampT sampling strategy
for the \estiT, \estiTT, and \estiTTF cases,
alongside experimental results.
\cite{ph2solidexp:79drie,ph2solidexp:80silv,ph2solidexp:83ishm,ph2solidexp:94mao}
The corresponding pressure-density curves calculated
from \sampTT simulation results
are nearly indistinguishable from those shown in the figure.
The \estiT curve overestimates the experimental results even at low densities,
which means that the two-body interaction on its own
is much too repulsive.
From the \estiTT curve,
we can see that including the attractive three-body interaction
decreases the pressure.
However,
the three-body interaction overcorrects the repulsion from the two-body interaction,
to the extent that
at the highest densities we see unphysical results where
the pressure decreases with density.
From the \estiTTF curve,
we see that including the four-body interactions in the estimate for the total energy
once again increases the pressure,
and drastically improves the agreement between simulation and experimental results
up to around $ 0.065 \icubang $.
As we move to higher densities,
the four-body interaction once again overestimates the pressure compared to the experimental data.

The effect seen here,
in which the inclusion of three-body interactions causes an underestimation
of the pressure with respect to the experimental data,
has also been seen in quantum Monte Carlo simulations of solid helium.
Barnes and Hinde found that
the combination of
the two-body Aziz potential
\cite{he4pes:87aziz}
and the three-body Cencek potential
\cite{he4pes:09cenc}
causes an underestimation of the pressure at high densities.
\cite{pathinteg:17barn_1}
Chang and Boninsegni,
performed PIMC simulations of solid helium
using two separate pairs of two-body and three-body interaction potentials,
\cite{he4pes:79aziz,hepes3b:73bruc,he4pes:97janz,hepes3b:96cohe}
and found in both cases that the simulation results underestimated the pressure.
\cite{pathinteg:01chan}

%%%%%%%%%%%%%%%%%%%%%%%%%%%%%%%%%%%%%%%%%%%%
%%% RADIAL PAIR DISTRIBUTION FUNCTION
%%%%%%%%%%%%%%%%%%%%%%%%%%%%%%%%%%%%%%%%%%%%

\subsection{Radial pair distribution function} \label{subsec:radial_pair_distribution}

The inclusion of higher-order many-body interactions during sampling
affects the structure of the solid.
To see this change,
we can look at the radial pair distribution function $ g(r) $,
whose (un-normalized) estimator in the simulation is given by
\begin{equation} \label{eq:radial_pair_distribution:gr_estimator}
    \hat{g}(r)
    =
    \left\langle 
        \sum_{n < m}^{N}
        \sum_{i}^{P}
        \delta (r - | \vb{r}_{i,n} - \vb{r}_{i,m} |)
    \right\rangle \, ,
\end{equation}
\noindent
where
$ \vb{r}_{i, n} $ is the position of the bead
at the timeslice with index $ i $ and
the particle with index $ n $.
We can also look at the centroid radial pair distribution function $ g_c(r) $,
whose (un-normalized) estimator is given by
\begin{equation} \label{eq:radial_pair_distribution:gcrc_estimator}
    \hat{g_c}(r)
    =
    \left\langle 
        \sum_{n < m}^{N}
        \delta (r - | \vb{c}_{n} - \vb{c}_{m} |)
    \right\rangle \, ,
\end{equation}
\noindent
where $ \vb{c}_n $ is the centroid of the particle with index $ n $,
given by
\begin{equation} \label{eq:radial_pair_distribution:centroid}
    \vb{c}_n
    =
    \frac{1}{P}
    \sum_{i = 1}^{P}
    \vb{r}_{i,n} \, .
\end{equation}

\begin{figure*}
    \centering
    \includegraphics[width=\linewidth]{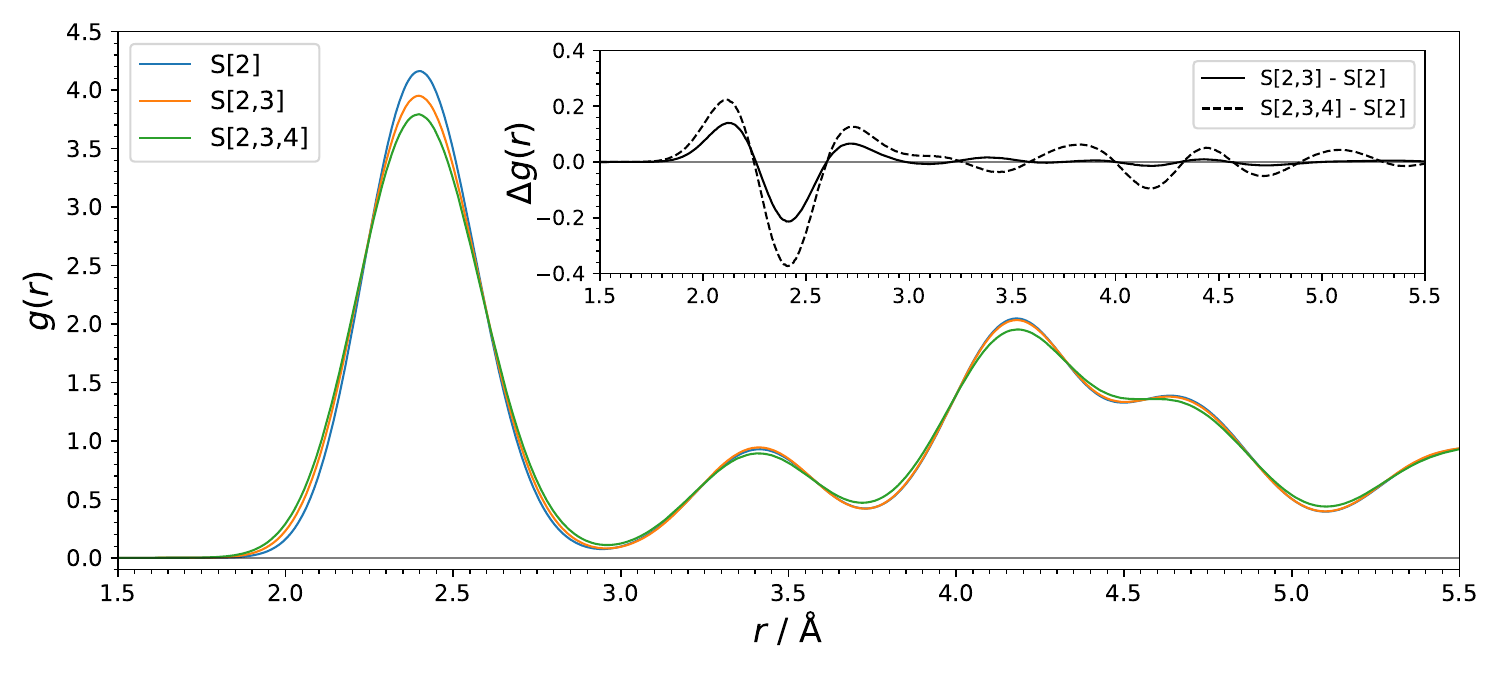}
    \caption{
        The radial pair distribution function $ g(r) $
        of solid parahydrogen at a density of $ \rho = 0.1 \icubang $,
        collected from simulations performed using the
        \sampT (blue),
        \sampTT (orange), and
        \sampTTF (green) sampling strategies.
        The estimates are done using Eq.~(\ref{eq:radial_pair_distribution:gr_estimator}).
        In the inset,
        we show the differences between the $ g(r) $ curves
        from the \sampTT and \sampT simulations (black, solid),
        and from the \sampTTF and \sampT simulations (black dashed).
    }
    \label{fig:radial_pair_distances_highest_density}
\end{figure*}

\begin{figure*}
    \centering
    \includegraphics[width=\linewidth]{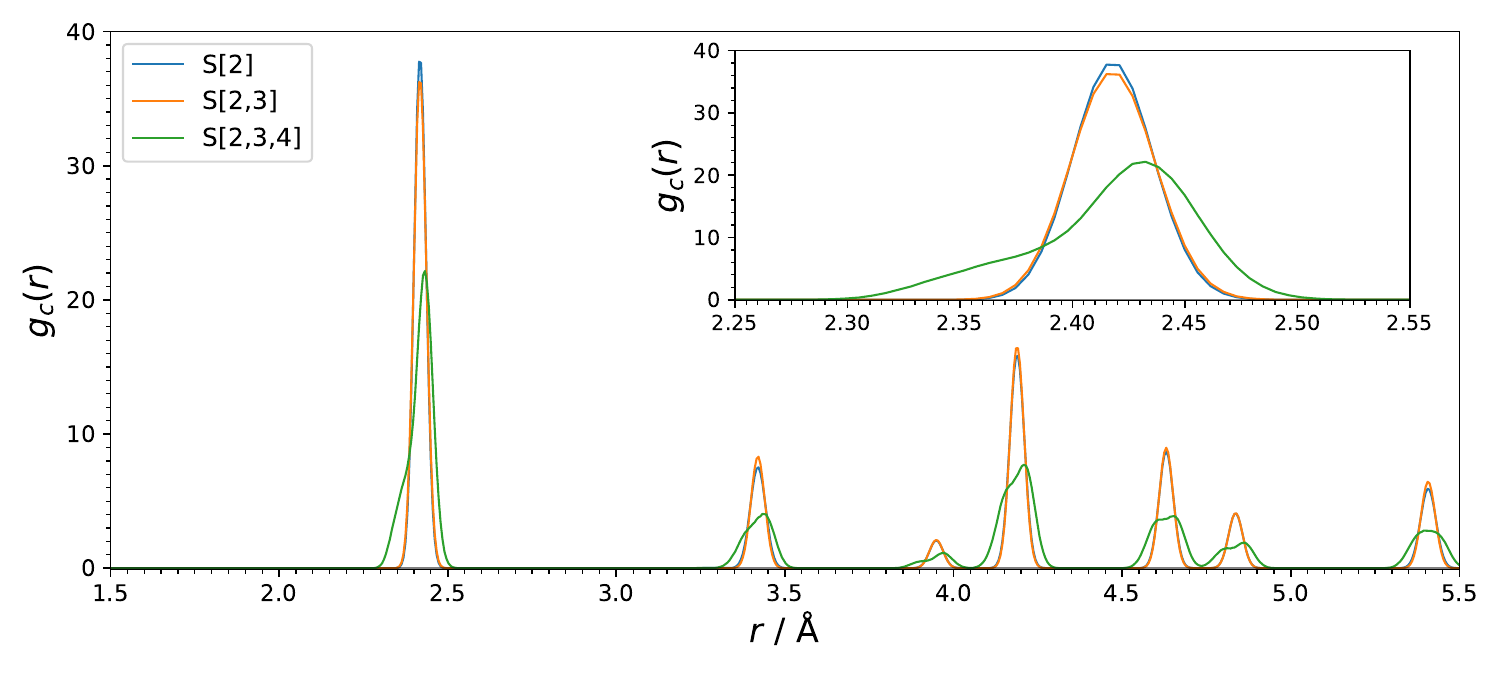}
    \caption{
        The centroid radial pair distribution function $ g_c(r) $
        of solid parahydrogen at a density of $ \rho = 0.1 \icubang $,
        collected from simulations performed using the
        \sampT (blue),
        \sampTT (orange), and
        \sampTTF (green) sampling strategies.
        The estimates are done using Eq.~(\ref{eq:radial_pair_distribution:gcrc_estimator}).
        In the inset,
        we zoom into the curve around the first shell.
    }
    \label{fig:centroid_radial_pair_distances_highest_density}
\end{figure*}

In Fig.~(\ref{fig:radial_pair_distances_highest_density}),
we show the normalized radial pair distribution function $ g(r) $
of the \sampT, \sampTT, and \sampTTF simulations at $ 0.1 \icubang $.
In Fig.~(\ref{fig:centroid_radial_pair_distances_highest_density}),
we show the normalized centroid radial pair distribution function $ g_c(r) $
under the same conditions. All three sampling strategies produce similar distributions.

In an earlier study,
\cite{pathinteg:22ibra}
we found that the distribution of the distances
from each bead to the centroid of their corresponding molecule,
\begin{equation} \label{eq:centroid_distribution}
    \hat{c}(r)
    =
    \left\langle 
        \sum_{n = 1}^{N}
        \sum_{i = 1}^{P}
        \delta (r - | \vb{r}_{i,n} - \vb{c}_{n} |)
    \right\rangle \, ,
\end{equation}
\noindent
is essentially unchanged between the \sampT and \sampTT simulations.
Our current simulations reproduce the same curves
as found in Fig.~(10) of Ref.~[\onlinecite{pathinteg:22ibra}].
Our estimate of Eq.~(\ref{eq:centroid_distribution})
for the \sampTTF simulations at $ 0.1 \icubang $
is the same as that from the \sampT and \sampTT simulations.
This indicates that the many-body interactions
do not change the overall quantum mechanical ``spreading'' of
each individual molecule.

The $ g(r) $ curve from the \sampTT simulation is slightly flatter and wider
than the $ g(r) $ curve from the \sampT simulation around the first shell.
From the second shell onwards,
the \sampT and \sampTT results are nearly indistinguishable.
The $ g_c(r) $ curve for the \sampTT simulation
is also slightly wider than that of the \sampT simulation around the first shell.
This indicates that the three-body potential softens the total interaction
in such a way that
the centroid of each molecule
has a greater translational displacement about its nominal lattice site,
but the quantum mechanical ``spreading'' of each molecule does not increase.

Given that the widening of the $ g(r) $ distribution
when moving from the \sampT case to the \sampTT case
is likely caused by the attractive interaction energy of the three-body PES,
it is surprising to see that the $ g(r) $ distribution widens
even further in the \sampTTF case,
when the strongly repulsive four-body PES is included.
This softening of the $ g(r) $ distribution is also pronounced beyond the first shell.
Looking at the $ g_c(r) $ curves
in the inset of Fig.~(\ref{fig:centroid_radial_pair_distances_highest_density}),
the cause becomes apparent.

The four-body PES used in this work
is a function of all six relative pair distances between the four molecules.
Although the relationship between the six side lengths and the interaction energy
is fairly complicated,
as a general rule,
we can decrease the four-body interaction energy by increasing the average side length.
\cite{fourbody:24ibra}
Consider a cluster of four \parahyd molecules,
one at each corner of a perfect tetrahedron.
We can lower the total interaction energy of this cluster
by moving two of the molecules slightly closer together,
and the other two molecules slightly farther apart,
in a way that causes the average side length of the cluster to increase.
This trend also applies to geometries found in the \hcp lattice
other than the perfect tetrahedron.
In other words,
we decrease the four-body interaction energy
by having certain pairs of molecules move slightly closer together,
and other pairs of molecules move slightly farther apart,
which causes the translational symmetry breaking in the \hcp lattice.
This is why including the four-body PES during sampling causes
the peaks in the $ g(r) $ curve to widen,
and the peaks in the $ g_c(r) $ curve to slightly split.

The translational symmetry breaking in the \hcp lattice
introduced by the \sampTTF simulations
is very likely to be a simulation artifact.
We should remind the readers that
these simulations are done at $ 0.1 \icubang $,
whereas Fig.~(\ref{fig:pressure_vs_density_eos_coarse_pert2b}) shows that
the simulations predict pressures that exceed the experimental results
beyond around $ 0.065 \icubang $.
At such high densities,
many-body interactions beyond the four-body interaction become significant.
It is possible that this symmetry breaking would vanish
if five-body and even higher-order many-body interactions
were included in these simulations at this density.

Like the \sampTTF simulations,
the predicted pressure-density relationship by the \sampTT simulations
deviate greatly from the experimental data at $ 0.1 \icubang $.
However,
the \sampTT simulations do not introduce a symmetry breaking effect,
despite also lacking higher-order many-body interactions.
A possible reason for this discrepancy might be the fact that
the four-body PES is a function of six pair distances,
whereas
the three-body PES is a function of only three pair distances.
Because the four-body PES has more degrees of freedom,
the molecules in the solid are able to deviate
from their ideal \hcp lattice positions
in a way that causes symmetry breaking
and decreases the total four-body interaction energy.
In contrast,
the three-body PES may not have enough degrees of freedom for this to occur,
causing the molecules to remain around their ideal \hcp lattice positions.

%%%%%%%%%%%%%%%%%%%%%%%%%%%%%%%%%%%%%%%%%%%%
%%% CONCLUSION
%%%%%%%%%%%%%%%%%%%%%%%%%%%%%%%%%%%%%%%%%%%%

\section{Conclusion} \label{sec:conclusion}

In this work,
we performed PIMC simulations of solid parahydrogen
using \abinitio two-body, three-body, and four-body PESs.
We found that simulations that use
the \abinitio two-body, three-body, and four-body interaction energy estimators
reproduce the experimental pressure-density results very well
below densities of about $ 0.065 \icubang $.
In particular,
the agreement between the simulation and experimental results
jumps drastically going from
the \estiTT to the \estiTTF case.
The three-body and four-body PESs
affect the energy-density relationship of solid parahydrogen
even at densities as low as $ 0.024 \icubang $.
The \estiTTF case gives a prediction of the equilibrium density
that is much closer to the experimental result
than either the \estiT and \estiTT cases.

We investigated the use of strategies that include or omit
the many-body interactions during sampling
as a way to improve the simulation performance.
When we switch from the \sampT to the \sampTT strategy,
each of the energy components experiences a shift,
but these changes cancel each other out in such a way that
the total interaction energy is nearly unchanged.
Compared to the \sampT simulation,
the \sampTT simulation gives an estimate of the average kinetic energy per molecule
at equilibrium density that is slightly closer to experiment.
Including the three-body PES during sampling causes very slight
changes in the structure of the solid.
At the relatively high density of $ 0.1 \icubang $,
further including the four-body PES introduces translational symmetry breaking
in the \hcp lattice.
This symmetry breaking is likely a simulation artifact,
and might be best described as a cautionary tale
against using many-body interactions
at densities where even higher-order many-body interactions are significant.

The most likely means of improving the simulation results at higher densities
would be to include five-body and higher-order many-body interaction energies.
Tian \etal performed a computational study on
effects of many-body interactions on the EOS of solid $ ^{4} $He,
up to the six-body interaction.
\cite{manybody:06tian}
They found that,
at high densities,
the even-parity (two-body, four-body, six-body) interactions are overall repulsive,
while the odd-parity (three-body and five-body) interactions are overall attractive.
Each successive many-body interaction
has a weaker overall contribution
to the total interaction energy.
Their model for the pressure-density EOS of solid helium
showed greater agreement with the experimental results
with each successive higher-order many-body interaction.
It is reasonable to assume that similar improvements can be seen in solid parahydrogen.

However,
there are a number of challenges with including \abinitio higher-order many-body interactions
in a non-additive fashion as has been done in this work.
The electronic structure calculations needed to create the training data
become much more expensive as more hydrogen molecules are included.
More drastic approximations may be required to create
the \abinitio five-body and higher-order many-body PESs.
For example,
one might ignore the Counterpoise correction,
use a lower-quality calculation method,
use a smaller basis set,
or focus only on certain geometries that contribute the most to the solid.
In addition,
the evaluation of the total interaction energy during the estimation step
becomes much more expensive.
It is likely that an entirely different approach to including the many-body interaction
energies will need to be explored.

The lack of further higher-order many-body interactions
cannot explain all the shortcomings of the simulation results.
For example,
the higher-order many-body interactions
are extremely weak at low densities,
and thus we cannot use their absence
to explain
the deviation between the predicted and experimental
kinetic energy per molecule
at the equilibrium density.
Another way to improve the results
would be to switch from using isotropic PESs,
to PESs that include more degrees of freedom.

%%%%%%%%%%%%%%%%%%%%%%%%%%%%%%%%%%%%%%%%%%%%
%%% SUPPLEMENTARY MATERIAL
%%%%%%%%%%%%%%%%%%%%%%%%%%%%%%%%%%%%%%%%%%%%

\section*{SUPPLEMENTARY MATERIAL}
Refer to the supplementary material for descriptions of
the fit parameters for the modified Birch equation of state,
the energy-density curves near equilibrium for the \sampT and \sampTT simulations,
and the energy-density data used to create the equations of state in this paper.

%%%%%%%%%%%%%%%%%%%%%%%%%%%%%%%%%%%%%%%%%%%%
%%% DATA AVAILABILITY
%%%%%%%%%%%%%%%%%%%%%%%%%%%%%%%%%%%%%%%%%%%%

\section*{DATA AVAILABILITY}
The data that supports the findings of this study are available within the article and its supplementary material.

%%%%%%%%%%%%%%%%%%%%%%%%%%%%%%%%%%%%%%%%%%%%
%%% ACKNOWLEDGEMENTS
%%%%%%%%%%%%%%%%%%%%%%%%%%%%%%%%%%%%%%%%%%%%

\section*{Acknowledgements}
The authors acknowledge
the Natural Sciences and Engineering Research Council (NSERC) of Canada (RGPIN-2016-04403),
the Ontario Ministry of Research and Innovation (MRI),
the Canada Research Chair program (950-231024),
and the Canada Foundation for Innovation (CFI) (project No. 35232).
A.~I. acknowledges the support of the NSERC of Canada (CGSD3-558762-2021).

\end{document}

%%%%%%%%%%%%%%%%%%%%%%%%%%%%%%%%%%%%%%%%%%%%
%%% TITLE, AUTHORSHIP, EMAIL, AFFILIATIONS
%%%%%%%%%%%%%%%%%%%%%%%%%%%%%%%%%%%%%%%%%%%%

\title{Supplementary Information for:
Path-integral Monte Carlo simulations of solid parahydrogen using two-body,
three-body, and four-body ab initio interaction potential energy surfaces
}

\author{Alexander Ibrahim}
\affiliation{
    Department of Physics and Astronomy,
    University of Waterloo,
    200 University Avenue West, Waterloo, Ontario N2L 3G1, Canada
}
\affiliation{
    Department of Chemistry,
    University of Waterloo,
    200 University Avenue West, Waterloo, Ontario N2L 3G1, Canada
}

\author{Pierre-Nicholas Roy}
\email{pnroy@uwaterloo.ca}
\affiliation{
    Department of Chemistry,
    University of Waterloo,
    200 University Avenue West, Waterloo, Ontario N2L 3G1, Canada
}

\maketitle

\section{Fit parameters for the modified Birch EOS} \label{sec:birch_eos_fit}

The average energy per particle as a function of density
is fit to a modified Birch equation of state,
given by
\begin{equation} \label{supp:eq:energy_density:modified_birch_eos}
    \epsilon
    =
    \epsilon_0
    -
    \frac{P_0}{d \rho_0}
    +
    \frac{1}{\rho_0}
    \sum_{n=1}^{4}
    \kappa_{n} d^{2n / 3} \, ,
\end{equation}
\noindent
where $ d = \rho / \rho_0 $ and $ \rho_0 = 0.0261 \icubang $.
We do this for the energy-density curves collected from simulations performed using
both the \sampT and \sampTT sampling strategies,
and using either of the \estiT, \estiTT, or \estiTTF estimations.
In Tables~(\ref{tab:supp:modified_birch_eos_parameters_s2})~and~(\ref{tab:supp:modified_birch_eos_parameters_s23}),
we show the fit parameters for the modified Birch EOS
from simulations performed with the \sampT and \sampTT strategies,
respectively.
The fit parameters were found using \texttt{scipy.optimize.curve\_fit()}
with \texttt{scipy} version \texttt{1.14.1}.

As a goodness-of-fit measure, we pick
\begin{equation} \label{eq:weighted_chi_squared}
    \chi^2
    =
    \frac{
        \sum_{i=1}^{N} w_i \left( \epsilon_i - \epsilon(\rho_i) \right)^2
    }{
        \sum_{i=1}^{N} w_i
    } \, .
\end{equation}
\noindent
where
$ w_i = 1 / \sigma_i^2 $
is the reciprocal of the standard error of the mean associated with the $ i $th data sample,
$ \epsilon_i $ is the mean of the $ i $th data sample,
and
$ \epsilon(\rho_i) $ is the energy predicted by Eq.~(\ref{supp:eq:energy_density:modified_birch_eos})
at the $ i $th density $ \rho_i $.
The value of $ \chi^2 $ for each fit equation
is given in the captions of
Tables~(\ref{tab:supp:modified_birch_eos_parameters_s2})~and~(\ref{tab:supp:modified_birch_eos_parameters_s23}).

\begin{table} [H]
    \caption{
        Parameters for Eq.~(\ref{supp:eq:energy_density:modified_birch_eos}),
        for the energy-density curves created from
        the \estiT, \estiTT, and \estiTTF estimates,
        collected from simulations using the \sampT sampling strategy.
        The parameter $ \epsilon_0 $ is in unit of $ \wvn $,
        with the remaining five parameters in units of $ \wvn \ang^{-3} $.
        The fits for the \estiT, \estiTT, and \estiTTF estimates
        have $ \chi^2 $ values of
        $ 0.002832 $,
        $ 0.009615 $, and
        $ 0.001693 $,
        respectively.
    }
    \begin{ruledtabular}
        \begin{tabular}{lccc}
                       & \estiT                        & \estiTT                       & \estiTTF                       \\
        $ \epsilon_0 $ & $ -1.56505681 \times 10^{3} $ & $ -1.26250369 \times 10^{2} $ & $  3.56249208 \times 10^{3} $ \\
        $ P_0        $ & $ -3.75230222 \times 10^{0} $ & $  3.15494256 \times 10^{0} $ & $  1.51704339 \times 10^{1} $ \\
        $ \kappa_1   $ & $  1.02906387 \times 10^{2} $ & $  4.71395809 \times 10^{1} $ & $ -1.66626893 \times 10^{2} $ \\
        $ \kappa_2   $ & $ -1.09622970 \times 10^{2} $ & $ -8.40506135 \times 10^{1} $ & $  1.28837072 \times 10^{2} $ \\
        $ \kappa_3   $ & $  4.63227491 \times 10^{1} $ & $  5.07239623 \times 10^{1} $ & $ -5.28651108 \times 10^{1} $ \\
        $ \kappa_4   $ & $ -4.15077483 \times 10^{0} $ & $ -9.01830199 \times 10^{0} $ & $  1.12007899 \times 10^{1} $ 
        \end{tabular}
    \end{ruledtabular}
    \label{tab:supp:modified_birch_eos_parameters_s2}
\end{table}

\begin{table} [H]
    \caption{
        Parameters for Eq.~(\ref{supp:eq:energy_density:modified_birch_eos}),
        for the energy-density curves created from
        the \estiT, \estiTT, and \estiTTF estimates,
        collected from simulations using the \sampTT sampling strategy.
        The units for the parameters match those given in the caption in
        Tab.~(\ref{tab:supp:modified_birch_eos_parameters_s2}).
        The fits for the \estiT, \estiTT, and \estiTTF estimates
        have $ \chi^2 $ values of
        $ 0.002102 $,
        $ 0.008917 $, and
        $ 0.002962 $,
        respectively.
    }
    \begin{ruledtabular}
        \begin{tabular}{lccc}
                       & \estiT                        & \estiTT                       & \estiTTF                      \\
        $ \epsilon_0 $ & $ -1.83424519 \times 10^{3} $ & $ -5.49258462 \times 10^{2} $ & $  2.54277684 \times 10^{3} $ \\
        $ P_0        $ & $ -5.05459078 \times 10^{0} $ & $  1.04887723 \times 10^{0} $ & $  1.05389673 \times 10^{1} $ \\
        $ \kappa_1   $ & $  1.14397738 \times 10^{2} $ & $  6.47249178 \times 10^{1} $ & $ -1.20773904 \times 10^{2} $ \\
        $ \kappa_2   $ & $ -1.17679352 \times 10^{2} $ & $ -9.60762919 \times 10^{1} $ & $  9.44807408 \times 10^{2} $ \\
        $ \kappa_3   $ & $  4.88952562 \times 10^{1} $ & $  5.45899350 \times 10^{1} $ & $ -4.07315282 \times 10^{1} $ \\
        $ \kappa_4   $ & $ -4.43279955 \times 10^{0} $ & $ -9.51174267 \times 10^{0} $ & $  9.55183896 \times 10^{0} $ 
        \end{tabular}
    \end{ruledtabular}
    \label{tab:supp:modified_birch_eos_parameters_s23}
\end{table}

\section{Energy-density curves near equilibrium} \label{sec:energy_density_equilibrium_s2_s23}

\begin{figure} [H]
    \centering
    \includegraphics[width=\linewidth]{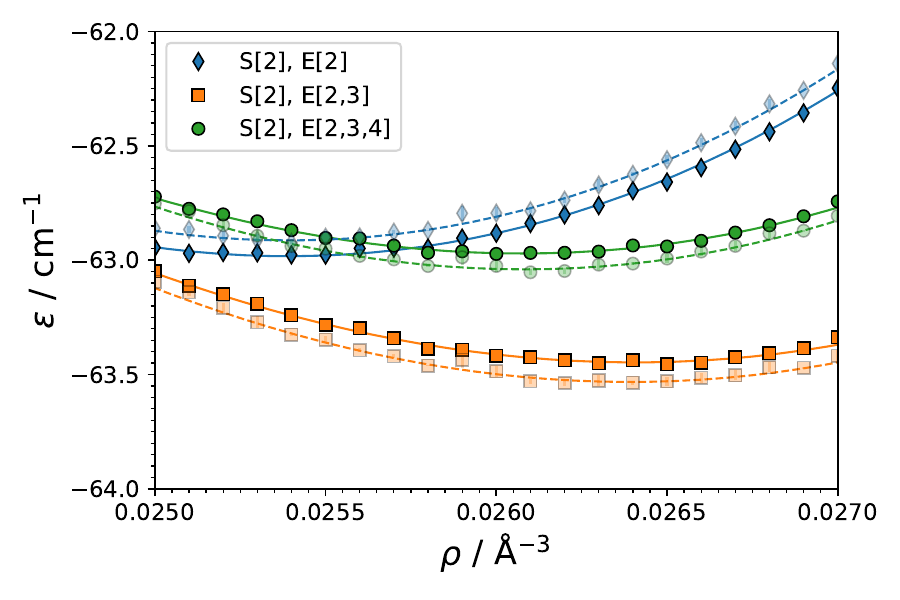}
    \caption{
        The energy per molecule $ \epsilon $ as a function of the density $ \rho $
        of solid parahydrogen around the equilibrium density
        for the
        \estiT (blue, diamonds),
        \estiTT (orange, squares),
        and \estiTTF (green, circles) estimates.
        Results from the \sampT simulations are shown
        with solid markers and solid lines,
        whereas results from the \sampTT simulations are shown
        with semitransparent markers
        and semitransparent dashed lines.
    }
    \label{fig:energy_vs_density_eos_equilibrium_pert2b_vs_pert2b3b}
\end{figure}

In Fig.~(\ref{fig:energy_vs_density_eos_equilibrium_pert2b_vs_pert2b3b}),
we show the energy-density curves around the equilibrium density
for the \sampT and \sampTT simulations.

\section{Energy-density data} \label{sec:energy_density_data}
As part of the supplementary information,
we have included the energy-density data used to create
the equations of state in the manuscript.
This includes the individual
kinetic,
two-body,
three-body, and
four-body interaction energies per particle,
as well as the two-body and three-body tail corrections.

The data is located within the submitted tarball.
Also in the tarball
is a \texttt{README.md} file with a more detailed description of the energy-density data.